\documentclass[aps,prl,reprint,groupedaddress,showpacs,showkeys]{revtex4-1}
\pdfoutput=1
\usepackage{graphicx}  
\usepackage{dcolumn}   
\usepackage{bm}        
\usepackage{amssymb}   
\usepackage{color}     
\def\da{$\Delta\alpha/\alpha$}

\begin{document}

\title{Indications of a spatial variation of the fine structure constant}
\author{J.~K.~Webb$^{1}$} \author{J.~A.~King$^{1}$} \author{M.~T.~Murphy$^{2}$}
\author{V.~V.~Flambaum$^{1}$} \author{R.~F.~Carswell$^{3}$}
\author{M.~B.~Bainbridge$^{1}$} 
\affiliation{$^{1}$School of Physics, University of New South Wales, Sydney, NSW 2052, Australia} 
\affiliation{$^{2}$Centre for Astrophysics and Supercomputing, Swinburne University of 
Technology, Mail H30, PO Box 218, Victoria 3122, Australia} 
\affiliation{$^{3}$Institute of Astronomy, Madingley Road, Cambridge, CB3 0HA, England.} 
\date{\today}

\begin{abstract} 
  We previously reported Keck telescope observations suggesting a
  smaller value of the fine structure constant, $\alpha$, at high
  redshift. New Very Large Telescope (VLT) data, probing a different
  direction in the universe, shows an inverse evolution; $\alpha$
  increases at high redshift. Although the pattern could be due to as
  yet undetected systematic effects, with the systematics as presently
  understood the combined dataset fits a spatial dipole, significant
  at the 4.2$\sigma$ level, in the direction right ascension $17.5 \pm
  0.9$ hours, declination $-58 \pm 9$ degrees. The independent VLT and
  Keck samples give consistent dipole directions and amplitudes, as do
  high and low redshift samples.  A search for systematics, using
  observations duplicated at both telescopes, reveals none so far
  which emulate this result.
\end{abstract}

\maketitle


{\it Quasar spectroscopy as a test of fundamental physics.---} The
vast light-travel times to distant quasars allow us to probe physics at high
redshift. The relative wavenumbers, $\omega_z$, of atomic
transitions detected at redshift $z=\lambda_{obs}/\lambda_{lab} -1$, can be compared
with laboratory values, $\omega_0$, via the relationship $\omega_z = \omega_0 +
Q\left(\alpha_z^2 - \alpha_0^2 \right)/\alpha_0^2 $ where the coefficient $Q$ measures
the sensitivity of a given transition to a change in $\alpha$. The variation in both
magnitude and sign of $Q$ for different transitions is a significant advantage of
the Many Multiplet method\,\cite{Webb99short,Dzuba99short}, helping to combat potential
systematics.

The first application of this method, 30 measurements of
$\Delta\alpha/\alpha=\left(\alpha_z-\alpha_0\right)/\alpha_0$, indicated a smaller
$\alpha$ at high redshift at the $3\sigma$ significance level. By 2004 we had made
143 measurements of $\alpha$ covering a wide redshift range, using further data from
the Keck telescope obtained by 3 separate groups, supporting our earlier findings,
that towards that general direction in the universe at least, $\alpha$ may have been
smaller at high redshift, at the $5\sigma$ 
level\,\cite{Webb01short,Murphy03short,Murphy04short}. 
The constant factor at that point was (undesirably) the telescope and spectrograph.


{\it New data from the VLT.---} We have now analysed a large dataset
from a different observatory, the VLT. Full details and searches for
systematic errors will be given elsewhere\cite{King11short, Koch11short}. Here we
summarize the evidence for spatial variation in $\alpha$ emerging from the
combined Keck+VLT samples. Quasar spectra, obtained from the ESO Science
Archive, were selected, prioritising primarily by expected signal to noise but
with some preference given to higher redshift objects and to objects giving
more extensive sky coverage. The ESO {\sc midas} pipeline was used for the first
data reduction step, including wavelength calibration, although enhancements
were made to derive a more robust and accurate wavelength solution from an
improved selection of thorium-argon calibration lamp emission
lines\,\cite{Murphy07short}. Echelle spectral orders from several exposures of a
given quasar were combined using {\sc uves\_popler}\,\cite{uves_popler}. A
total of 60 quasar spectra from the VLT have been used for the present work, 
yielding 153 absorption systems. Absorption systems were identified via a careful
visual search of each spectrum, using {\sc rdgen}\,\cite{rdgen}, scanning for
commonly detected transitions at the same redshift, hence aligned in velocity
coordinates. Several transition matches were required for acceptance and, given
the high spectral resolution, chance matches were eliminated.


{\it Absorption system modelling.---} As in our previous studies, {\sc vpfit} was
used to model the profiles in each absorption system\,\cite{vpfit} with some
enhancements, described in\,\cite{King11short}. A comprehensive list of the transitions
used, their laboratory wavelengths, oscillator strengths, and $Q$ coefficients are
compiled in\,\cite{Murphy03short, King11short}.

The following general procedures were adhered to:
{\it (i)} For each absorption system, physically related parameters (redshifts
and $b$-parameters) are tied, in order to minimise the required number of free
parameters and derive the strongest possible constraints on line positions,
and hence {\da}. 
{\it (ii)} Parameters were tied only for species with similar ionisation potentials,
to minimise possible introduction of random effects on $\alpha$, mimicked by
spatial (and hence velocity) segregation effects; 
{\it (iii)} Line broadening is typically dominated by turbulent rather than
thermal motion.  Both limiting-case models were applied and {\da} determined 
for each. The final {\da} was derived from a likelihood-weighted average;
{\it (iv)} Where appropriate and available, isotopic shifts and hyperfine 
structure are included in the fitting procedure;
{\it (v)} Velocity structures were determined initially choosing the strongest
unsaturated transitions in each system. Normalised residuals across each
transition fitted were examined and the fit progressively refined with the
introduction of each additional transition to the fit;
{\it (vi)} Transitions falling in spectral regions contaminated by telluric
features or atmospheric absorption were discarded. Any data regions
contaminated by cosmic rays, faulty CCD pixels, or any other unidentified
noise effects, were also discarded; 
{\it (vii)} A few gravitational lenses were identified by being difficult or
impossible to model successfuly. The non-point source quasar image and the resultant
complex line-of-sight geometry can significantly alter apparent relative line
strengths. These systems were discarded;
{\it (viii)} In all cases we derived the final model without solving
for {\da}.  The introduction of {\da} as an additional free parameter 
was only done once the profile velocity structure had been finalised,
eliminating any possible bias towards a `preferred' {\da}. 
One potential consequence of this approach might conceivably be a small
bias on {\da} {\it towards} zero, should some `fitting-away' of {\da}
occur by column density adjustments or velocity structure decisions.
The reverse is not true, i.e. it cannot bias towards a non-zero {\da}.
For details of all the points above see\,\cite{King11short}.

{\sc vpfit}\,\cite{vpfit} minimises $\chi^2$ simultaneously over all species.
Whilst the strongest components may appear in all species, weaker components can
sometimes fall below the detection threshold and hence are excluded, such that a
component which appears in MgII, for example, does not appear in FeII. There is no
solution to this (known) problem but its effect merely adds an additional random
scatter on {\da} for an ensemble of observations.


{\it Spatially dependent $\alpha$.---} An initial inspection of {\da} vs
redshift for the new VLT dataset reveals a redshift trend, opposite in sign
compared to the earlier Keck data. Splitting each sample at $z=1.8$, our 2004
Keck sample\,\cite{Murphy04short} gave 
$\langle${\da}$\rangle_{ z<1.8} = -0.54 \pm 0.12 \times 10^{-5}$ and 
$\langle${\da}$\rangle_{ z>1.8} = -0.74 \pm 0.17 \times 10^{-5}$. 
The present VLT sample, discussed in detail in\,\cite{King11short}, gives
$\langle${\da}$\rangle_{ z<1.8} = -0.06 \pm 0.16 \times 10^{-5}$ and
$\langle${\da}$\rangle_{ z>1.8} = +0.61 \pm 0.20 \times 10^{-5}$.
Errors here and throughout this paper are 1$\sigma$ estimates.
Our VLT result above for $z<1.8$ agrees with the VLT data presented 
in \cite{Srianand07short}.

Errors on individual {\da} values for our VLT sample are $\sigma^2_{tot} =
\sigma^2_{stat} + \sigma^2_{rand}$, where $\sigma^2_{rand}$ was derived
empirically by fitting a constant {\da} to the sample, i.e. monopole--only,
using a modification of the Least Trimmed Squares (LTS) method, where only
85\% of data, those points with the smallest squared residuals, are fitted.
$\sigma_{rand}$ was assumed constant for all absorbers and found to be
$\approx 0.9 \times 10^{-5}$, showing that the scatter in the VLT {\da} is
greater than expected on the basis of statistical-errors alone. Errors on
{\da} for the Keck sample are discussed in\,\cite{Murphy03short}, although we
derive a new estimate of $\sigma_{rand}=1.74$ for the Keck points using the
LTS method, again relative to a monopole--only fit to the Keck sample.

The Keck (Mauna Kea, Hawaii) and VLT (Paranal, Chile) locations on Earth are
separated by $45^{\circ}$ in latitude and hence, on average, observe different
directions on the sky. The $\langle${\da}$\rangle$ results above suggest exploring
a simple spatial dependence using the combined dataset.

The Keck sample we use is as presented in \cite{Murphy04short} with
minor modifications: 3 points were removed. 2 had been included erroneously
(from a spectrum known to have calibration problems) and 1 further point
was clipped, having a residual greater than 3$\sigma$ against 
a modified LTS fit to the Keck data.

We fit 3 different models to the 
3 datasets (i.e. Keck, VLT and combined samples).  Initially we try a 
dipole+monopole model, {\da}$\,=A\cos\Theta + m$, where $m$ allows an 
offset from the terrestrial value, $\Theta$ is the angle on the sky between
quasar sightline and best-fit dipole position, and $A$ is the dipole amplitude.
Noting the theoretical interpretation of the monopole term is unclear,
we fit a second model, without the monpole, {\da}$\,=A\cos\Theta$. 
Thirdly, in order to explore a possible spatial gradient in $\alpha$, 
we assign a distance to each {\da} measurement of $r(z)=ct(z)$ where $c$ is the speed 
of light and $t(z)$ is the look-back time at redshift $z$. The model is then
{\da}$\,=Ar\cos\Theta$.

To estimate the dipole significance we bootstrap the sample, repeatedly
randomising the association between {\da} and the absorption system location
in space (i.e. quasar sightline and absorption redshift). A dipole is fitted
and its $\chi^2$ derived at each realisation, to obtain a $\chi^2$ probability
distribution. This gives the probability of fitting a dipole to the data and
obtaining a value of $\chi^2$ less than or equal to that observed for the real
sample by chance alone, i.e. the statistical significance of a dipole model
compared to a monopole, and hence an uncertainty estimate for the dipole
amplitude, $A$.

All 3 models give a detection significance in the range $4.1 - 4.2 \sigma$ and the 
best-fit parameters and associated errors (given in the figure captions) vary only 
slightly. Figure 1 illustrates an all-sky map for the best-fit no-monopole model, 
using equatorial co-ordinates. Approximate 1$\sigma$ error contours are derived from 
the covariance matrix. Figure 2 illustrates the {\da} binned data and the best-fit 
dipole+monopole model. Figure 3 illustrates {\da} vs look-back time
distance projected onto the dipole axis, $r\cos{\Theta}$, using the best-fit dipole
parameters for this model. This model seems to represent the data reasonably
well and {\da} appears distance-dependent, the correlation being significant 
at the $4.2\sigma$ level.

An alternative to the LTS method described above, to allow for any unknown
additional contribution to the errors on individual {\da} measurements,
one can assume $\sigma^2_{tot} = \sigma^2_{stat}$ and
iteratively trim the sample during model fitting. This provides a further 
test of whether the apparent spatial gradient in $\alpha$ is dominated by a subset of the
data, perhaps more prone to some unknown systematic than the remainder. Adopting
$\sigma^2_{tot} = \sigma^2_{stat}$ will tend to result in higher significance
levels. Figure 4 illustrates this test and shows that the apparent dipole seems
robust to data trimming.
\begin{figure}
\includegraphics[viewport=0 0 649 378,width=85mm]{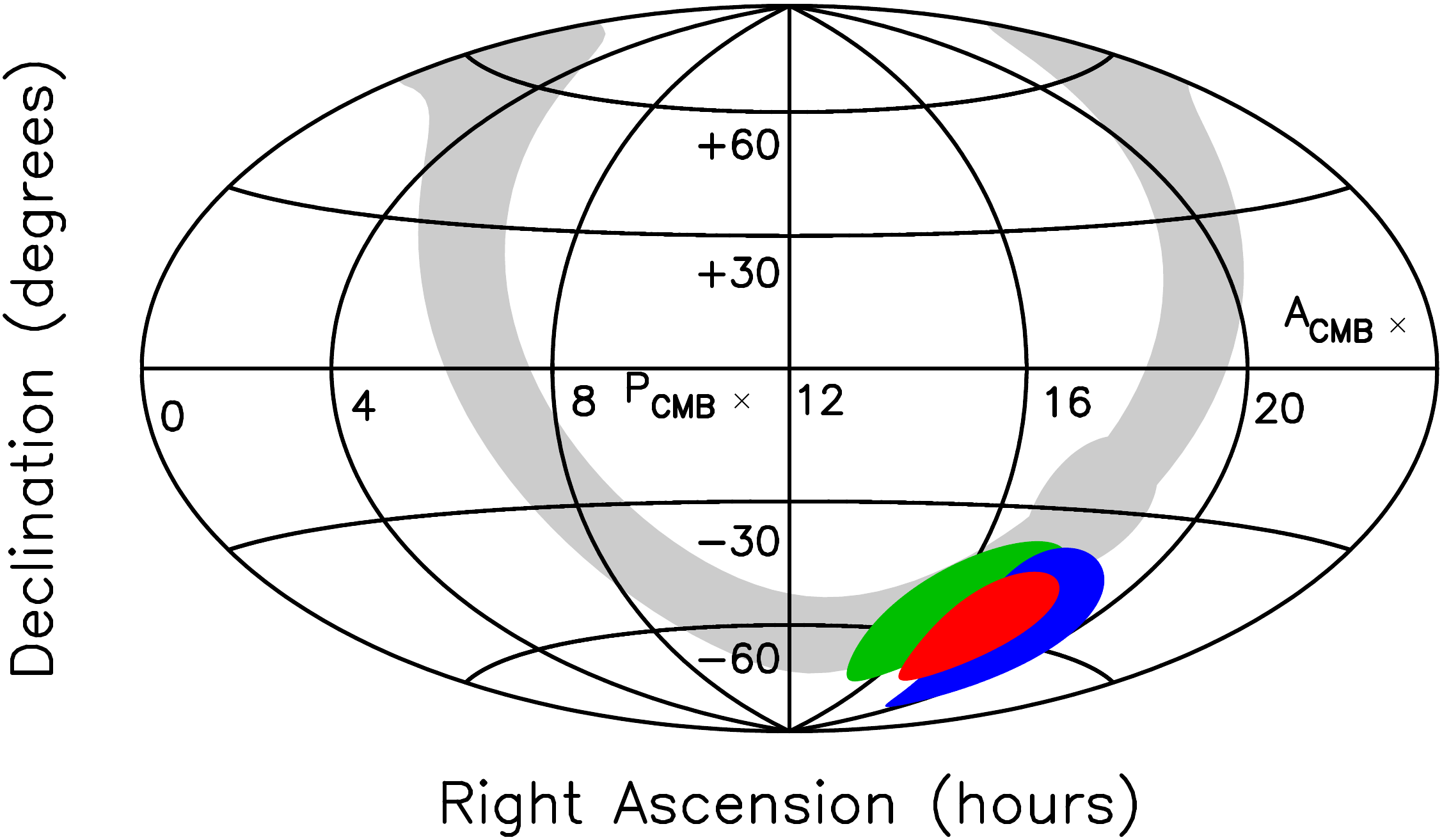}
\caption{\label{fig1} All-sky plot in equatorial coordinates showing the
independent Keck (green, leftmost) and VLT (blue, rightmost) best-fit dipoles, and the combined
sample (red, centre), for the dipole model, {\da}$\,=A\cos\Theta$, with
$A=(1.02 \pm 0.21) \times 10^{-5}$. Approximate 1$\sigma$ confidence contours
are from the covariance matrix. The best-fit dipole is at right ascension
$17.4 \pm 0.9$ hours, declination $-58 \pm 9$ degrees and is statistically preferred
over a monopole-only model at the $4.1\sigma$ level. For this model, a bootstrap
analysis shows the chance-probability of the dipole aligments being as good or
closer than observed is 6\%. For a dipole+monopole model this increases to 14\%.
The cosmic microwave background dipole and antipole are illustrated for comparison.
}

\end{figure}

\begin{figure}
\includegraphics[viewport=0 0 651 480, width=85mm]{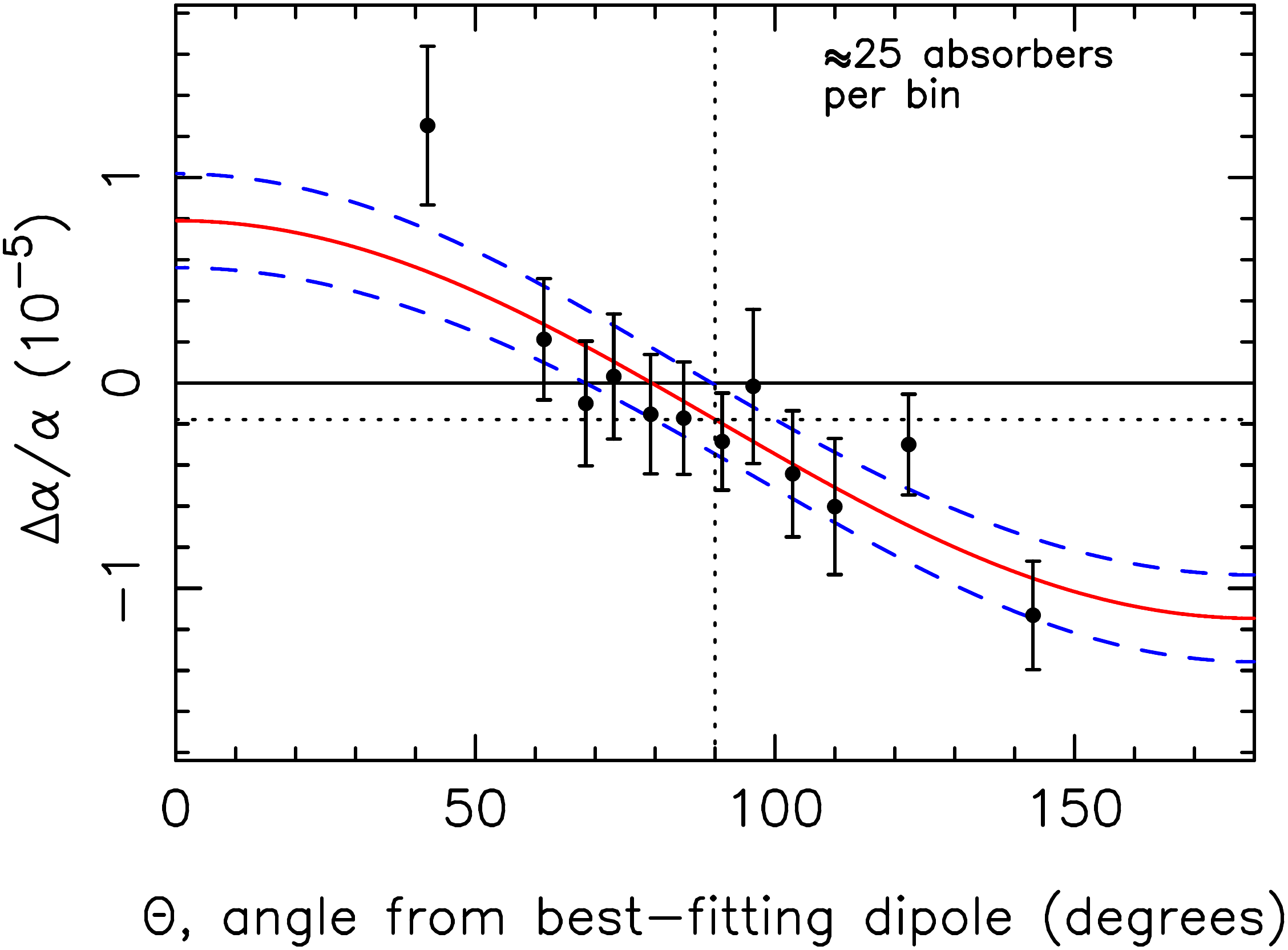}
\caption{\label{fig2}
{\da} for the combined Keck and VLT data vs angle $\Theta$ from the best-fit
dipole position (best-fit parameters given in Figure 1 caption).  Dashed lines 
illustrate $\pm 1\sigma$ errors. For a discussion on the monopole term, 
see\,\cite{King11short}. 
}
\end{figure}

\begin{figure} 
\includegraphics[viewport=0 0 638 479,width=85mm]{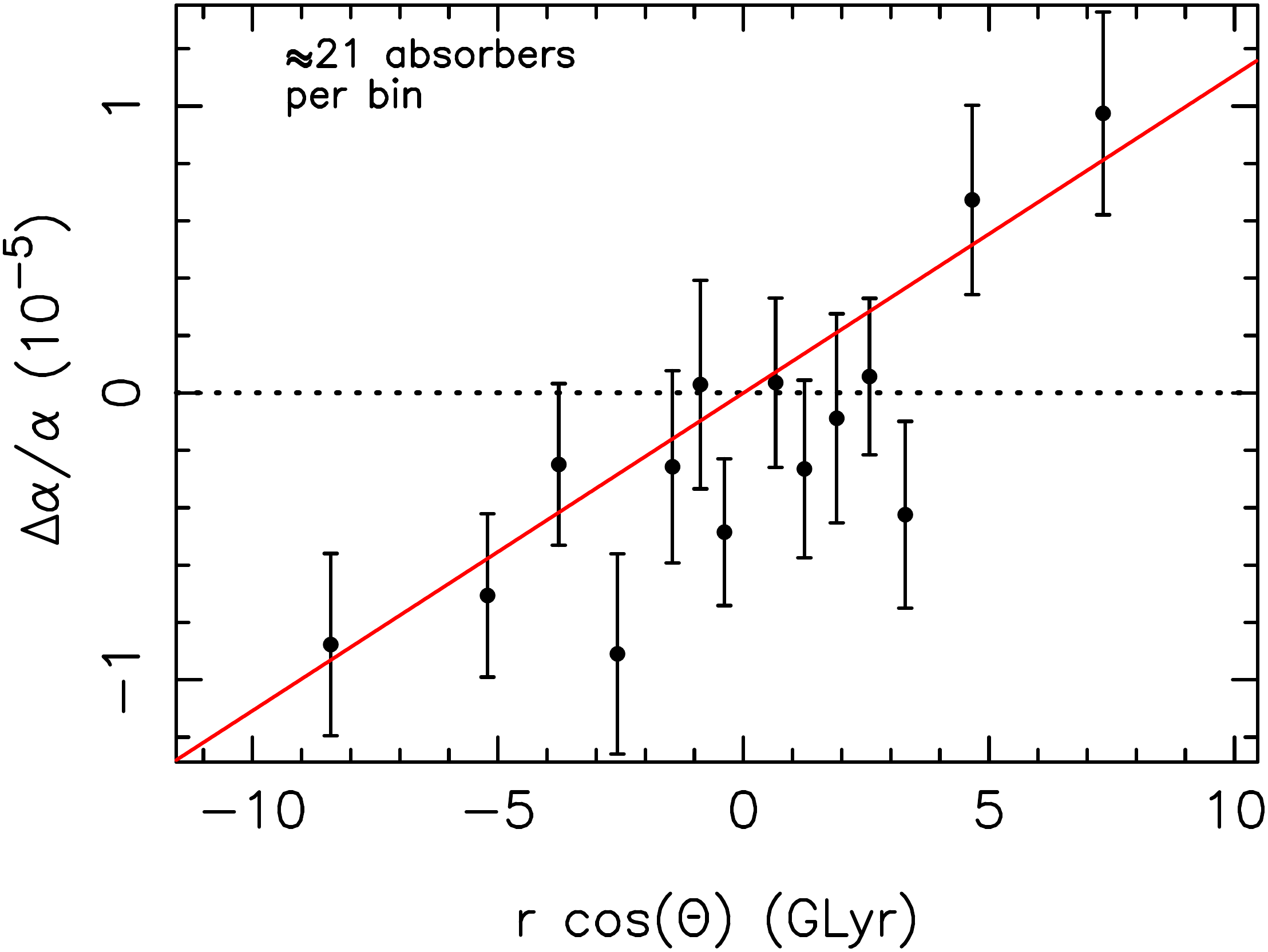}
\caption{\label{fig3} 
{\da} vs $Ar \cos\Theta$ showing an apparent gradient in $\alpha$ along the 
best-fit dipole. The best-fit direction is at right ascension $17.5 \pm 0.9$ 
hours, declination $-58 \pm 9$ degrees, for which 
$A=(1.1 \pm 0.25) \times 10^{-6}$ GLyr$^{-1}$. A spatial gradient is statistically 
preferred over a monopole-only model at the $4.2\sigma$ level. 
A cosmology with parameters
$\left( H_0, \Omega_M, \Omega_{\Lambda} \right) = \left( 70.5, 0.2736, 0.726
\right)$ was used\,\cite{Hinshaw09short}.} 
\end{figure}

\begin{figure}
\includegraphics[viewport=0 0 783 512,width=85mm]{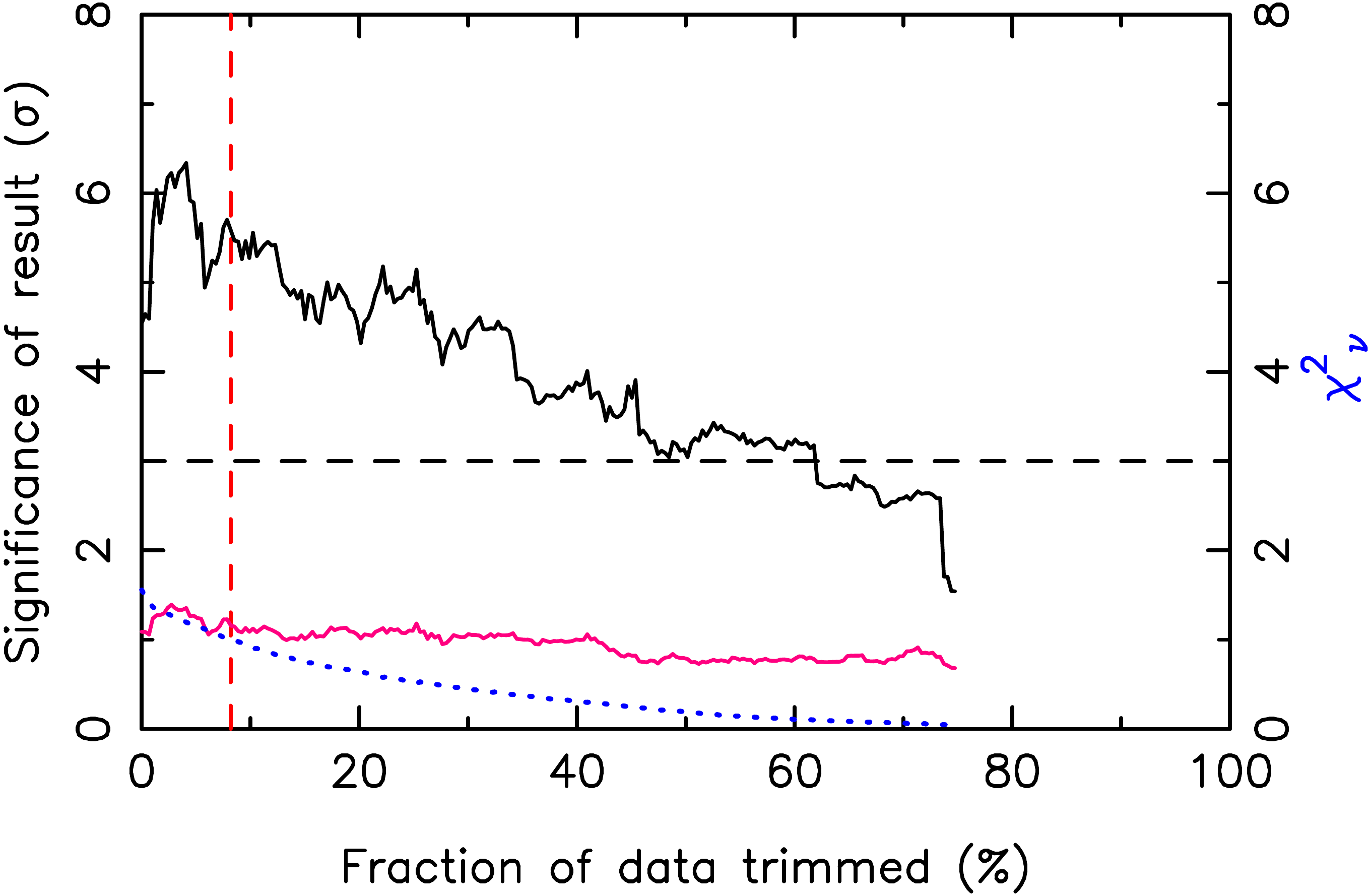}
\caption{\label{fig4} As an alternative to increasing {\da} error bars, to
account for the additional scatter in the data as described in the text, we 
instead use $\sigma^2_{tot} = \sigma^2_{stat}$ and iteratively clip the most 
deviant {\da} value, fitting {\da}$\,=Ar\cos\Theta$. Approximately 60\% of 
the data must be discarded before the significance drops below $3\sigma$ 
showing the dipole signal is not due to a small subset of the data. The solid 
(pink) line at the bottom of the graph shows the dipole amplitude in units of 
$10^{-6}$ Glyr$^{-1}$. The dotted (blue) line at the bottom of the graph shows 
$\chi^2_{\nu}$ and the vertical dashed (red) line illustrates $\chi^2_{\nu} = 1$ 
when $\sim 8$\% of the data has been trimmed, at which point the significance 
is $\sim5.5\sigma$ }
\end{figure}


{\it Empirical test for systematics.---} One potential systematic in the data
could arise if there were slight mechanical mis-alignments of the
slits for the 2 arms of the UVES spectrograph on the VLT. This could cause
wavelength shifts between spectral features falling in the blue and red
arms.  However, this specific effect appears to be substantially
smaller than required to explain values of {\da}$\sim 10^{-5}$ seen in the
present work \cite{Molaro08short}.

A more subtle but related effect may be slight off-centre placement of the
quasar image in the spectrograph slit, by different amounts for different
exposures, at different wavelength settings. This may apply to either or
both Keck and VLT spectra. Since spectrograph slit illuminations are
different for quasar (point source) and ThAr calibration lamp (uniform
illumination), the subsequent combination of individual exposures to form a
1-dimensional spectrum may then contain relative velocity shifts between
spectral segments coming from different exposures.  This effect will exist in
our data at some level and it is clearly important to know the impact on an
ensemble of measurements of $\alpha$.

Fortunately, 6 quasars in our sample have both Keck and VLT spectra,
allowing a direct and empirical check on the effect above, and indeed any
other systematics which produce relative velocity shifts along the spectrum.
To do this we selected small spectral segments, each a few \AA~ wide, flanked
by unabsorbed continuum flux from the quasar, and fitted Voigt profiles using
{\sc vpfit}, but adding an additional free parameter allowing a velocity shift
between the Keck and VLT segments, $\delta v(\lambda_{obs})_i$, where
$\lambda_{obs}$ is the observed wavelength and $i$ refers to the $i^{th}$
quasar. All available absorption lines in the 6 spectra were used, including
both Lyman-$\alpha$ forest lines and heavy element lines but excluding
telluric features. In this way we can map any effective relative distortions
in the calibrations between each pair of spectra. A total of 694 measurements
were used from the 6 pairs of spectra over the observed wavelength range 
$3506 < \lambda < 8945$\AA.

We formed a composite function $\delta v(\lambda_{obs})$ after first normalising
$\langle \delta v(\lambda_{obs})_i\rangle = 0$ for each $i$ 
to remove any potential small constant velocity offsets from each spectrum
(expected from off-centering of the quasar in the spectrograph slit), which
cannot influence $\alpha$.

Finally, we fit the composite $\delta v(\lambda_{obs})$ with a linear function
$f(\delta v) = a \lambda_{obs} + b$ where $a = (-7 \pm 14) \times 10^{-5}\,
\mathrm{km\,s^{-1}\AA^{-1}}$, $b = 0.38
\pm 0.71 \ \mathrm{km\,s^{-1}}$. The final $f(\delta v)$ thus shows a weak 
(but statistically
insignificant) velocity drift, and provides an empirical transformation
between the Keck and VLT wavelength scales. For each VLT quasar absorption
system, we modify the input laboratory wavelengths used in the Voigt profile
fitting procedure $\lambda_{lab}$ to $\lambda^\prime_{lab} = \lambda_{lab} +
\Delta\lambda_{lab}$ where $\Delta\lambda_{lab} = \lambda_{lab}
\,\delta v\,(\lambda_{obs}) / c $, and finally use the $\lambda^\prime_{lab}$ to
re-compute {\da} for the entire sample.

There was one complicating aspect of this effect excluded from the discussion
above, arising from a 7$^{th}$ spectral pair.  The $\delta v(\lambda_{obs})_7$
showed a more significant non-zero slope than the other 6, suggesting a small
but significant calibration problem with that particular spectrum. We
therefore applied a slightly more complicated transformation to the data to
allow for this, using a Monte Carlo simulation to estimate the potential
impact on our full combined Keck and VLT sample of both the previous
effect measured in 6 quasars {\it plus} the effect derived from the 7$^{th}$
quasar simultaneously, applied in appropriate proportions. The full details of
this analysis will be discussed separately in\,\cite{Koch11short}.

A systematic of the same magnitude as that from the 7$^{th}$ pair cannot
be present in any large fraction of our data, otherwise it would generate
large numbers of noticeable outliers. If we apply $f(\delta v)$ from the 6 quasar
pairs, the significance of the dipole+monopole model {\da}$\,=A\cos\Theta + m$,
is reduced to $3.1\sigma$. Blindly including
the effect of the 7$^{th}$ pair under a Monte Carlo method reduces the significance
to a most likely value of $2.2\sigma$. However, in this circumstance we introduce
significant extra scatter into the data above that 
already observed, implying that it over-estimates any systematic effect
of this type. Additionally, the trend of $\delta
v(\lambda_{obs})_i$ against wavelength is different in magnitude and sign
for each quasar pair, implying that these effects are likely to average
out for an ensemble of observations. Thus, application of the effect 
as described above should be regarded as extreme in terms of impact
on estimating {\da}.


{\it Conclusions.---} Quasar spectra obtained using 2
separate observatories show a spatial variation in the relative spacings
of absorption lines which could be due to an as yet undetected systematic 
effect, or a dipole variation of $\alpha$. A fit to the dipole gives a
significance of $\gtrsim 4.2\sigma$, assuming the error bars described above.
Assuming a dipole interpretation, the two datasets exhibit internal 
consistency and the directions of the independently derived spatial dipoles
agree. The magnitudes of the apparent {\da} variation in both datasets also agree.
A subset of the quasar spectra observed at both observatories permits a direct 
test for systematics.  So far, none are found which are likely to emulate the
apparent cosmological dipole in $\alpha$ we detect. Consistency with other 
astronomical data is discussed in\,\cite{BerengutPRD11}.
Consistency with laboratory, meteorite, and Oklo natural reactor is
discussed in\,\cite{Berengut10}.  Short-wavelength oscillatory 
variations in the wavelength scale such as those reported by\,\cite{Griest10short},
\,\cite{Whitmore10}, 
do not significantly impact on our results. To explain our results in terms
of systematics would require at least 2 different, finely tuned, effects. Future
similar measurements targeting the apparent pole and anti-pole directions will
maximise detection sensitivity, and further observations duplicated on 2 independent
telescopes will better constrain systematics. Most importantly, an independent 
technique is required to check these results.  Qualitatively, our results 
could violate the equivalence principle and infer a very large or infinite
universe, within which our `local' Hubble volume represents a tiny fraction, 
with correspondingly small variations in the physical constants.

This work is supported by the Australian Research Council. We thank Steve
Curran, Elliott Koch and Julian Berengut for discussions throughout this work.


\begin{thebibliography}{10}%
\makeatletter
\providecommand \@ifxundefined [1]{%
 \ifx #1\undefined \expandafter \@firstoftwo
 \else \expandafter \@secondoftwo
\fi
}%
\providecommand \@ifnum [1]{%
 \ifnum #1\expandafter \@firstoftwo
 \else \expandafter \@secondoftwo
\fi
}%
\providecommand \enquote [1]{``#1''}%
\providecommand \bibnamefont  [1]{#1}%
\providecommand \bibfnamefont [1]{#1}%
\providecommand \citenamefont [1]{#1}%
\providecommand\href[0]{\@sanitize\@href}%
\providecommand\@href[1]{\endgroup\@@startlink{#1}\endgroup\@@href}%
\providecommand\@@href[1]{#1\@@endlink}%
\providecommand \@sanitize [0]{\begingroup\catcode`\&12\catcode`\#12\relax}%
\@ifxundefined \pdfoutput {\@firstoftwo}{%
 \@ifnum{\z@=\pdfoutput}{\@firstoftwo}{\@secondoftwo}%
}{%
 \providecommand\@@startlink[1]{\leavevmode}%
 \providecommand\@@endlink[0]{}%
}{%
 \providecommand\@@startlink[1]{%
  \leavevmode
  \pdfstartlink
   attr{/Border[0 0 1 ]/H/I/C[0 1 1]}%
   user{/Subtype/Link/A<</Type/Action/S/URI/URI(#1)>>}%
  \relax
 }%
 \providecommand\@@endlink[0]{\pdfendlink}%
}%
\providecommand \url  [0]{\begingroup\@sanitize \@url }%
\providecommand \@url [1]{\endgroup\@href {#1}{\urlprefix}}%
\providecommand \urlprefix [0]{URL }%
\providecommand \Eprint[0]{\href }%
\@ifxundefined \urlstyle {%
  \providecommand \doi [1]{doi:\discretionary{}{}{}#1}%
}{%
  \providecommand \doi [0]{doi:\discretionary{}{}{}\begingroup
  \urlstyle{rm}\Url }%
}%
\providecommand \doibase [0]{http://dx.doi.org/}%
\providecommand \Doi[1]{\href{\doibase#1}}%
\providecommand \bibAnnote [3]{%
  \BibitemShut{#1}%
  \begin{quotation}\noindent
    \textsc{Key:}\ #2\\\textsc{Annotation:}\ #3%
  \end{quotation}%
}%
\providecommand \bibAnnoteFile [2]{%
  \IfFileExists{#2}{\bibAnnote {#1} {#2} {\input{#2}}}{}%
}%
\providecommand \typeout [0]{\immediate \write \m@ne }%
\providecommand \selectlanguage [0]{\@gobble}%
\providecommand \bibinfo [0]{\@secondoftwo}%
\providecommand \bibfield [0]{\@secondoftwo}%
\providecommand \translation [1]{[#1]}%
\providecommand \BibitemOpen[0]{}%
\providecommand \bibitemStop [0]{}%
\providecommand \bibitemNoStop [0]{.\EOS\space}%
\providecommand \EOS [0]{\spacefactor3000\relax}%
\providecommand \BibitemShut [1]{\csname bibitem#1\endcsname}%
\bibitem{Webb99short}%
  \BibitemOpen
  \bibfield{author}{%
  \bibinfo {author} {\bibfnamefont{J.~K.}\ \bibnamefont{{Webb}}}
  \emph{et~al.},\ }%
  \bibfield{journal}{%
  \Doi{10.1103/PhysRevLett.82.884}{\bibinfo {journal} {Phys.~Rev.~Lett.}}\ }%
  \textbf{\bibinfo {volume} {82}},\ \bibinfo {pages} {884} (\bibinfo {year}
  {1999})%
  \bibAnnoteFile{NoStop}{Webb99short}%
\bibitem{Dzuba99short}%
  \BibitemOpen
  \bibfield{author}{%
  \bibinfo {author} {\bibfnamefont{V.~A.}\ \bibnamefont{{Dzuba}}}, \bibinfo
  {author} {\bibfnamefont{V.~V.}\ \bibnamefont{{Flambaum}}},\ and\ \bibinfo
  {author} {\bibfnamefont{J.~K.}\ \bibnamefont{{Webb}}},\ }%
  \bibfield{journal}{%
  \Doi{10.1103/PhysRevLett.82.888}{\bibinfo {journal} {Phys.~Rev.~Lett.}}\ }%
  \textbf{\bibinfo {volume} {82}},\ \bibinfo {pages} {888} (\bibinfo {year}
  {1999})%
  \bibAnnoteFile{NoStop}{Dzuba99short}%
\bibitem{Webb01short}%
  \BibitemOpen
  \bibfield{author}{%
  \bibinfo {author} {\bibfnamefont{J.~K.}\ \bibnamefont{{Webb}}}
  \emph{et~al.},\ }%
  \bibfield{journal}{%
  \Doi{10.1103/PhysRevLett.87.091301}{\bibinfo {journal} {Phys.~Rev.~Lett.}}\
  }%
  \textbf{\bibinfo {volume} {87}},\ \bibinfo {pages} {091301} (\bibinfo {year}
  {2001})%
  \bibAnnoteFile{NoStop}{Webb01short}%
\bibitem{Murphy03short}%
  \BibitemOpen
  \bibfield{author}{%
  \bibinfo {author} {\bibfnamefont{M.~T.}\ \bibnamefont{{Murphy}}}, \bibinfo
  {author} {\bibfnamefont{J.~K.}\ \bibnamefont{{Webb}}},\ and\ \bibinfo
  {author} {\bibfnamefont{V.~V.}\ \bibnamefont{{Flambaum}}},\ }%
  \bibfield{journal}{%
  \Doi{10.1046/j.1365-8711.2003.06970.x}{\bibinfo {journal}
  {Mon.~Not.~Roy.~Astron.~Soc.}}\ }%
  \textbf{\bibinfo {volume} {345}},\ \bibinfo {pages} {609} (\bibinfo {year}
  {2003})%
  \bibAnnoteFile{NoStop}{Murphy03short}%
\bibitem{Murphy04short}%
  \BibitemOpen
  \bibfield{author}{%
  \bibinfo {author} {\bibfnamefont{M.~T.}\ \bibnamefont{{Murphy}}}, \bibinfo
  {author} {\bibfnamefont{V.~V.}\ \bibnamefont{{Flambaum}}},\ and\ \bibinfo
  {author} {\bibfnamefont{J.~K.}\ \bibnamefont{{Webb}}},\ }%
  in\ \emph{\bibinfo {booktitle} {Astrophysics, Clocks and Fundamental Constants}},\ \bibinfo
  {series} {Lecture Notes in Physics (Springer, Berlin, Heidelberg, New York, 2004)}, Vol.\ \bibinfo {volume} {648},\ \bibinfo
  pp.\ \bibinfo {pages} {131--150}%
  \bibAnnoteFile{NoStop}{Murphy04short}%
\bibitem{King11short}%
  \BibitemOpen
  \bibfield{author}{%
  \bibinfo {author} {\bibfnamefont{J.~A.}\ \bibnamefont{{King}}}
  \emph{et~al.},\ }%
  \enquote{\bibinfo {title} {{Spatial variation in the fine-structure constant
  -- new results from VLT/UVES}},}\  
\bibinfo {note} {to be published.}%
  \bibAnnoteFile{Stop}{King11short}%
\bibitem{Koch11short}%
  \BibitemOpen
  \bibfield{author}{%
  \bibinfo {author} {\bibfnamefont{F.~E.}\ \bibnamefont{{Koch}}}
  \emph{et~al.},\ }%
  \enquote{\bibinfo {title} {{Spatial variation in the fine-structure constant
  -- a search for systematic effects}},}\  
\bibinfo {note} {to be published.}%
  \bibAnnoteFile{Stop}{Koch11short}%
\bibitem{Murphy07short}%
  \BibitemOpen
  \bibfield{author}{%
  \bibinfo {author} {\bibfnamefont{M.~T.}\ \bibnamefont{{Murphy}}}
  \emph{et~al.},\ }%
  \bibfield{journal}{%
  \Doi{10.1111/j.1365-2966.2007.11768.x}{\bibinfo {journal}
  {Mon.~Not.~Roy.~Astron.~Soc.}}\ }%
  \textbf{\bibinfo {volume} {378}},\ \bibinfo {pages} {221} (\bibinfo {year}
  {2007})%
  \bibAnnoteFile{NoStop}{Murphy07short}%
\bibitem{uves_popler}%
  \BibitemOpen
  \bibfield{author}{%
  \bibinfo {author} {\bibfnamefont{M.~T.}\ \bibnamefont{{Murphy}}},\ }%
  \enquote{\bibinfo {title} {{\sc uves\_popler}},}\  (\bibinfo {year} {2010}),\
  \bibinfo {note} {\url{http://astronomy.swin.edu.au/~mmurphy/UVES_popler}}%
  \bibAnnoteFile{NoStop}{uves_popler}%
\bibitem{rdgen}%
  \BibitemOpen
  \bibfield{author}{%
  \bibinfo {author} {\bibfnamefont{R.~F.}\ \bibnamefont{{Carswell}}},\ }%
  \enquote{\bibinfo {title} {{\sc rdgen}},}\  (\bibinfo {year} {2004}),\
  \bibinfo {note} {\url{http://www.ast.cam.ac.uk/~rfc/rdgen.html}}%
  \bibAnnoteFile{NoStop}{rdgen}%
\bibitem{vpfit}%
  \BibitemOpen
  \bibfield{author}{%
  \bibinfo {author} {\bibfnamefont{R.~F.}\ \bibnamefont{{Carswell}}}\ and\
  \bibinfo {author} {\bibfnamefont{J.~K.}\ \bibnamefont{{Webb}}},\ }%
  \enquote{\bibinfo {title} {{{\sc vpfit} - Voigt profile fitting program.
  Version 9.5}},}\  (\bibinfo {year} {2010}),\ \bibinfo {note}
  {\url{http://www.ast.cam.ac.uk/~rfc/vpfit.html}}%
  \bibAnnoteFile{NoStop}{vpfit}%
\bibitem{Srianand07short}%
  \BibitemOpen
  \bibfield{author}{%
  \bibinfo {author} {\bibfnamefont{R.}~\bibnamefont{{Srianand}}}, \bibinfo
  {author} {\bibfnamefont{H.}~\bibnamefont{{Chand}}}, \bibinfo {author}
  {\bibfnamefont{P.}~\bibnamefont{{Petitjean}}},\ and\ \bibinfo {author}
  {\bibfnamefont{B.}~\bibnamefont{{Aracil}}},\ }%
  \bibfield{journal}{%
  \Doi{10.1103/PhysRevLett.99.239002}{\bibinfo {journal} {Phys.~Rev.~Lett.}}\
  }%
  \textbf{\bibinfo {volume} {99}},\ \bibinfo {pages} {239002} (\bibinfo {year}
  {2007})%
  \bibAnnoteFile{NoStop}{Srianand07short}%
\bibitem{Molaro08short}%
  \BibitemOpen
  \bibfield{author}{%
  \bibinfo {author} {\bibfnamefont{P.}~\bibnamefont{{Molaro}}} \emph{et~al.},\
  }%
  \bibfield{journal}{%
  \Doi{10.1051/0004-6361:20078864}{\bibinfo {journal} {Astron.~Astrophys.}}\ }%
  \textbf{\bibinfo {volume} {481}},\ \bibinfo {pages} {559} (\bibinfo {year}
  {2008})%
  \bibAnnoteFile{NoStop}{Molaro08short}%
\bibitem{BerengutPRD11}%
 \BibitemOpen
  \bibfield{author}{%
  \bibinfo {author} {\bibfnamefont{J.~C.}~\bibnamefont{{Berengut}}}, \bibinfo
  {author} {\bibfnamefont{V.~V.}~\bibnamefont{{Flambaum}}}, \bibinfo {author}
  {\bibfnamefont{J.~A.}~\bibnamefont{{King}}}, \bibinfo {author}
  {\bibfnamefont{S.~J.}~\bibnamefont{{Curran}}},\ and\ \bibinfo {author}
  {\bibfnamefont{J.~K.}~\bibnamefont{{Webb}}},\ }%
  \bibfield{journal}{%
  {\bibinfo {journal} {Phys.~Rev.~D}}\
  }%
  \textbf{\bibinfo {volume} {83}},\ \bibinfo {pages} {123506} (\bibinfo {year}
  {2011})%
  \bibAnnoteFile{NoStop}{BerengutPRD11}%
\bibitem{Berengut10}%
  \BibitemOpen
  \bibfield{author}{%
  \bibinfo {author} {\bibfnamefont{J.~C.}\ \bibnamefont{{Berengut}}}\ and\
  \bibinfo {author} {\bibfnamefont{V.~V.}\ \bibnamefont{{Flambaum}}},\ }%
  \Eprint{http://arxiv.org/abs/1008.3957}{arXiv:1008.3957}%
  \bibAnnoteFile{NoStop}{Berengut10}%
\bibitem{Griest10short}%
  \BibitemOpen
  \bibfield{author}{%
  \bibinfo {author} {\bibfnamefont{K.}~\bibnamefont{{Griest}}}, \bibinfo
  {author} {\bibfnamefont{J.~B.}\ \bibnamefont{{Whitmore}}}, \bibinfo {author}
  {\bibfnamefont{A.~M.}\ \bibnamefont{{Wolfe}}}, \bibinfo {author}
  {\bibfnamefont{J.~X.}\ \bibnamefont{{Prochaska}}}, \bibinfo {author}
  {\bibfnamefont{J.~C.}\ \bibnamefont{{Howk}}},\ and\ \bibinfo {author}
  {\bibfnamefont{G.~W.}\ \bibnamefont{{Marcy}}},\ }%
  \bibfield{journal}{%
  \Doi{10.1088/0004-637X/708/1/158}{\bibinfo {journal} {Astrophys.~J.}}\ }%
  \textbf{\bibinfo {volume} {708}},\ \bibinfo {pages} {158} (\bibinfo {year}
  {2010})%
  \bibAnnoteFile{NoStop}{Griest10short}%
\bibitem{Whitmore10}%
  \BibitemOpen
  \bibfield{author}{%
  \bibinfo {author} {\bibfnamefont{J.~B.}\ \bibnamefont{{Whitmore}}}, \bibinfo
  {author} {\bibfnamefont{M.~T.}\ \bibnamefont{{Murphy}}},\ and\ \bibinfo
  {author} {\bibfnamefont{K.}~\bibnamefont{{Griest}}},\ }%
  \bibfield{journal}{%
  \Doi{10.1088/0004-637X/723/1/89}{\bibinfo {journal} {Astrophys.~J.}}\ }%
  \textbf{\bibinfo {volume} {723}},\ \bibinfo {pages} {89}
 (\bibinfo {year} {2010}),\
  \bibAnnoteFile{NoStop}{Whitmore10}%
\bibitem{Hinshaw09short}%
  \BibitemOpen
  \bibfield{author}{%
  \bibinfo {author} {\bibfnamefont{G.}~\bibnamefont{{Hinshaw}}} \emph{et~al.},\
  }%
  \bibfield{journal}{%
  \Doi{10.1088/0067-0049/180/2/225}{\bibinfo {journal} {Astrophys.~J.~Supp.~Ser.}}\
  }%
  \textbf{\bibinfo {volume} {180}},\ \bibinfo {pages} {225} (\bibinfo {year}
  {2009})%
  \bibAnnoteFile{NoStop}{Hinshaw09short}%
\end{thebibliography}
%

\vskip -0.5cm
\begin{figure}[!ht]
\begin{minipage}{\textwidth}
\includegraphics[viewport=41 65 468 728,angle=-90,width=160mm]{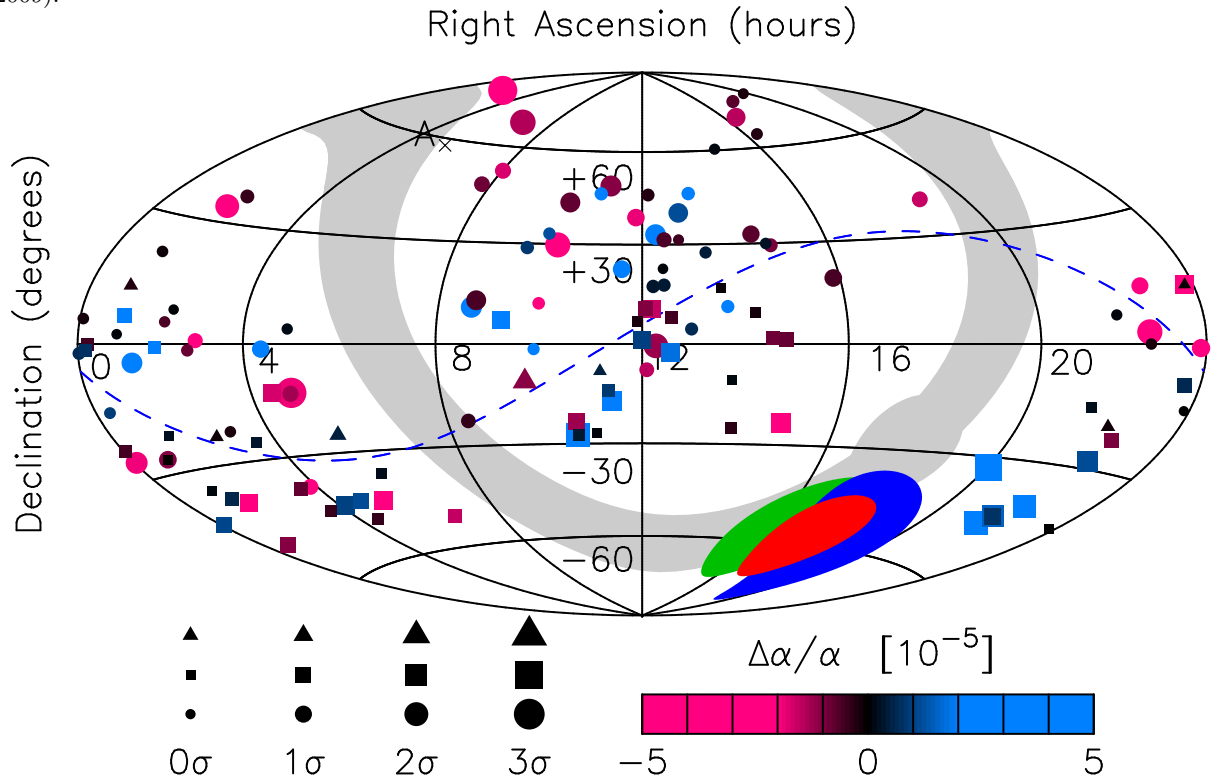}
\vskip 0cm
\caption{\label{fig5} Supplementary figure. Same all-sky illustration
  as in Fig.~1 showing the combined Keck and VLT {\da}
  measurements. Squares are VLT points. Circles are Keck
  points. Triangles are quasars observed at both Keck and VLT. Symbol
  size indicates deviation of {\da} from zero,
  i.e.~{\da}$\,=A\cos\Theta$. The blue dashed line shows the
  equatorial region. The grey shaded area shows the Galactic plane
  with the Galactic centre indicated as a bulge.  More and larger blue
  squares are seen closer to the $\alpha$-pole (red filled area) and
  more and larger red circles are seen closer to the
  $\alpha$-antipole.}
\end{minipage}
\end{figure}

\end{document}